\documentclass[preprint,amsmath,amssymb,prb,floatfix,superscriptaddress]{revtex4}

\usepackage[dvips]{graphicx}
\usepackage{dcolumn}
\usepackage{bm}
\usepackage{mathrsfs}

\begin{document}
\title{Magnetic exchange interaction between rare-earth and Mn ions in multiferroic hexagonal manganites}
\author{D. Talbayev}
   \email{diyar@lanl.gov}
   \affiliation{Center for Integrated Nanotechnologies, MS K771, Los Alamos National Laboratory, Los Alamos, NM 87545, USA}
\author{A. D. LaForge}
   \affiliation{Department of Physics, University of California San Diego, 9500 Gilman Drive, La Jolla, CA 92093, USA}
\author{S. A. Trugman} 
   \affiliation{Center for Integrated Nanotechnologies, MS K771, Los Alamos National Laboratory, Los Alamos, NM 87545, USA}
\author{N. Hur} 
   \affiliation{Department of Physics, Inha University, Incheon 402-751, Korea}
\author{A. J. Taylor} 
   \affiliation{Center for Integrated Nanotechnologies, MS K771, Los Alamos National Laboratory, Los Alamos, NM 87545, USA}
\author{R. D. Averitt}
   \affiliation{Center for Integrated Nanotechnologies, MS K771, Los Alamos National Laboratory, Los Alamos, NM 87545, USA}
   \affiliation{Department of Physics, Boston University, 590 Commonwealth Avenue, Boston, MA 02215, USA}
\author{D. N. Basov}
   \affiliation{Department of Physics, University of California San Diego, 9500 Gilman Drive, La Jolla, CA 92093, USA}
\date{\today}

\newcommand{\cm}{\:\mathrm{cm}^{-1}}
\newcommand{\T}{\:\mathrm{T}}
\newcommand{\mc}{\:\mu\mathrm{m}}
\newcommand{\ve}{\varepsilon}
\newcommand{\dg}{^\mathtt{o}}

\begin{abstract}
We report a study of magnetic dynamics in multiferroic hexagonal manganite HoMnO$_3$ by far-infrared spectroscopy. Low-temperature magnetic excitation spectrum of HoMnO$_3$ consists of magnetic-dipole transitions of Ho ions within the crystal-field split $J=8$ manifold and of the triangular antiferromagnetic resonance of Mn ions. We determine the effective spin Hamiltonian for the Ho ion ground state. The magnetic-field splitting of the Mn antiferromagnetic resonance allows us to measure the magnetic exchange coupling between the rare-earth and Mn ions.
\end{abstract}

\maketitle


Superexchange is one of the main concepts of magnetism in solids. Recently, superexchange was found to mediate magnetoelectric (ME) coupling in multiferroics - materials with coexisting magnetic and ferroelectric orders\cite{hur:392,lottermoser:541,cheong:13}. ME coupling may find important technological applications, for example, in an electric-write magnetic-read memory element\cite{scott:954}. Multiferroic hexagonal manganites display robust room-temperature ferroelectricity and remarkable ME behavior\cite{fiebig:818} that allow technological exploitation\cite{fujimura:1011}. In these multiferroics, the superexchange interaction between the rare-earth and Mn ions is responsible for the strong ME coupling that allows the control of ferromagnetism by an electric field\cite{lottermoser:541}. The details of the rare-earth/Mn superexchange coupling have heretofore remained unknown. Here, we report the measurement of the rare-earth/Mn superexchange in multiferroic hexagonal HoMnO$_3$ via the detection of an antiferromagnetic resonance by far-infrared (far-IR) spectroscopy. These results demonstrate the ferromagnetic nature of the rare-earth/Mn exchange that enables the electric-field control of magnetism in HoMnO$_3$.

HoMnO$_3$ (HMO) crystallizes in a hexagonal lattice, space group $P6_3cm$. The crystal structure consists of layers of corner-sharing trigonal MnO$_5$ bipyramids separated by layers of Ho$^{3+}$ ions\cite{lottermoser:541}. Ferroelectric polarization along the $c$ axis ($T_{FE}=875$ K) results from electrostatic and size effects that lead to the buckling of MnO$_5$ bipyramids and the displacement of the Ho$^{3+}$ ions out of the $a$-$b$ plane\cite{lottermoser:541,vanaken:164}. The magnetic structure is formed by Mn$^{3+}$ ions in a two-dimensional triangular network in the $a$-$b$ plane coupled by antiferromagnetic (AF) exchange and also by Ho$^{3+}$ ($^{5}I_8$) ions located at Ho(1) and Ho(2) lattice sites with point symmetries $C_{3v}$ and $C_3$, respectively\cite{munoz:1497, iliev:2488}. The AF exchange between Mn$^{3+}$ magnetic moments leads to their 120$\dg$ ordering at $T_N=72$ K\cite{vajk:87601,hur:3289}, and easy plane anisotropy confines the Mn spins to the $a$-$b$ plane. Two in-plane spin reorientation transitions occur at temperatures $T_{SR1}$$\approx$$38$ K and $T_{SR2}$$\approx$$5$ K associated with changes of magnetic symmetry to $P6'_3cm'$ and to $P6_3cm$, respectively\cite{vajk:87601,munoz:1497,fiebig:5620}. The spin reorientation at $T_{SR2}$$\approx$$5$ K is accompanied by antiferromagnetic ordering of the Ho$^{3+}$ ions located at the Ho(2) lattice sites\cite{munoz:1497}. Several manifestations of ME coupling in HMO have been documented, one of the most prominent being the presence of the reentrant phase with a strong magnetodielectric response\cite{lottermoser:541, lottermoser:220407, lorenz:87204, delacruz:60407,yen:180407}.

\begin{figure}[ht]
\begin{center}
\includegraphics[width=5in]{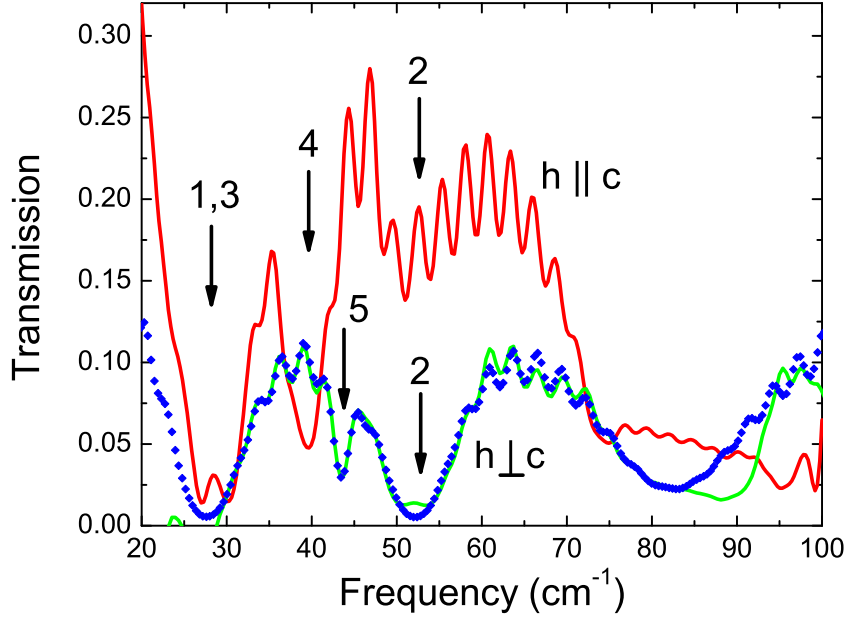}
\caption{\label{fig:lowtspectra}Power transmission spectra of HoMnO$_3$ with polarized far-IR light incident along the crystal's [110] direction at T=10 K. Arrows indicate magnetic absorption lines. The dotted line illustrates the simulation of the transmission spectrum using Eqs. (\ref{eq:epsilon}) and (\ref{eq:mu}) and the parameters $\epsilon_0=9.07$, $\sigma=0.9\cm$, with magnetic resonances centered at $27.5$, $43.5$, and $52.1\cm$.}
\end{center}
\end{figure}

The HMO single crystals were grown using an optical floating zone furnace\cite{hur:3289}. The material's magnetic phase diagram\cite{hur:3289} at temperatures above 5 K displays the reentrant phase discovered by Lorenz $et$ $al$\cite{lorenz:87204}. Polarized far-IR transmission of a 0.4 mm thick, (110) oriented crystal was measured as a function of temperature and applied magnetic field using a Bruker 66 Fourier-transform spectrometer coupled to an 8 Tesla split-coil superconducting magnet\cite{padilla:4710}. Static magnetic field $B$ was applied both parallel and perpendicular to the $c$ axis.

Figure~\ref{fig:lowtspectra} shows far-IR power transmission spectra of the HMO crystal at low temperatures and zero magnetic field for two polarizations of the light incident along the [110] direction - the magnetic $h$-field of the lightwave parallel ($h||c$) and perpendicular ($h\bot c$) to the $c$ axis. The spectra in Fig.~\ref{fig:lowtspectra} are normalized to the transmission through an empty aperture of the same size as the sample. The fringes with the period of $\approx3\cm$ seen in the spectra are due to multiple reflections of light within the sample. The transmission minima indicated by arrows correspond to magnetic resonance frequencies. The broad absorption in both polarizations at frequencies above $70\cm$ is of non-magnetic origin, as it exhibits no magnetic field dependence. The dotted line in Fig.~\ref{fig:lowtspectra} shows the simulated transmission, assuming that the magnetic and dielectric response of HMO consists of a collection of lorentzian oscillators:
\begin{eqnarray}
\label{eq:epsilon}
\epsilon(\omega) & = & \epsilon_0+i4\pi\sigma/\omega+\frac{f_0}{\omega_0^2-\omega^2-i\gamma_0\omega},\\
\mu(\omega)      & = & 1+\sum_j\frac{f_j}{\omega_j^2-\omega^2-i\gamma_j\omega},
\label{eq:mu}
\end{eqnarray}
where $\omega_j$, $\gamma_j$, and $f_j$ are frequency, relaxation rate, and oscillator strength of magnetic resonances. $\epsilon_0$ and $\sigma$ are constants, and $\omega_0$, $\gamma_0$, and $f_0$ are introduced to account for the non-magnetic absorption above 70 cm$^{-1}$.

\begin{figure}[ht]
\begin{center}
\includegraphics[width=5in]{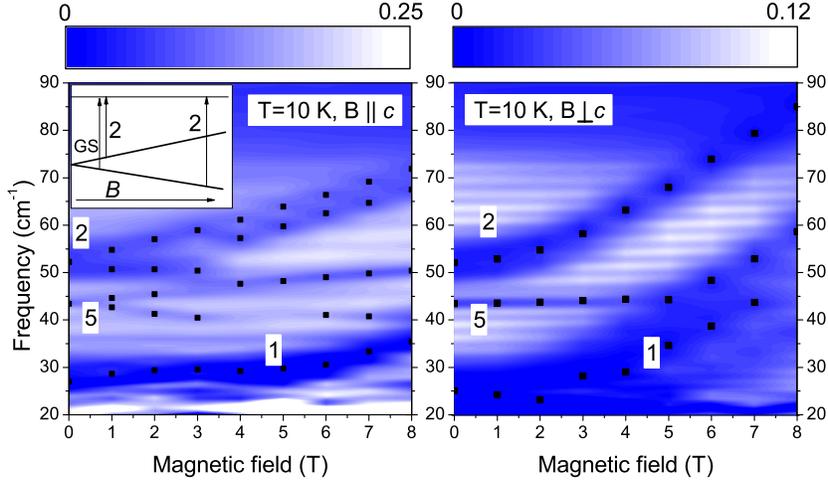}
\caption{\label{fig:hperpc}Transmission maps for the ($h\bot c$) polarization where darker color indicates lower transmission. Left panel - static field $B||c$, right panel - $B\bot c$. The bars above the panels set the absolute transmission scale. Symbols represent magnetic resonance positions extracted from fitting the measured transmission using Eqs. (\ref{eq:epsilon}) and (\ref{eq:mu}). Resonances are labeled by the same numbers used in Fig.~\ref{fig:lowtspectra}. The inset of the left panel is a scheme of the transition 2, which happens from the ground state (GS) doublet. The final state of the transition is assumed to be a singlet. In low magnetic field, the GS doublet splits and two resonance lines are observed as illustrated by the two vertical arrows on the left. In high magnetic field, the spitting of the GS becomes large enough for a thermal depopulation of the upper branch. Only one resonance line is observed, as the single vertical arrow on the right illustrates.}
\end{center}
\end{figure}

Color-coded transmission maps in Fig.~\ref{fig:hperpc} illustrate the magnetic field dependence of the resonance frequencies at T=10 K. To create the maps, we recorded the transmission spectra in 1 T steps between 0 and 8 T. The darker color in Fig.~\ref{fig:hperpc} indicates lower transmission and the white color indicates where the transmission is the highest; the symbols specify the positions of magnetic resonances extracted from the numerical simulation of the far-IR transmission. In the remainder of this Letter, we show that the magnetic field dependence of the resonance frequencies carries an unambiguous signature of the superexchange interaction between magnetic Ho$^{3+}$ and Mn$^{3+}$ ions that allows us to quantify the interaction.

We attribute the absorption lines labeled 1-4 in Figs.~\ref{fig:lowtspectra} and \ref{fig:hperpc} to crystal field excitations of Ho$^{3+}$ ions. The crystal fields of $C_{3v}$ and $C_3$ symmetry have qualitatively similar effects on the Ho ion. According to the work of Elliott and Stevens\cite{abragam}, the crystal field potential $V_{CF}$ can be written in terms of spin operators $J$, $J_z$, and $J_\pm$ for the manifold of states of constant $J$. The diagonal part of $V_{CF}$ depends only on $J$ and $J_z$. The off-diagonal part of $V_{CF}$ has non-zero matrix elements only between the states $m$ and $m'$ that differ by $\Delta m=6$, $m'=m\pm6$, in the field of $C_3$ symmetry and by $\Delta m=3,6$, $m'=m\pm3,6$, in the field of $C_{3v}$ symmetry. In zero applied magnetic field, the degenerate $J=8$ manifold splits into a set of singlet and doublet states of the form $\left|\pm 6,\pm 3,0 \right\rangle$ and $\left|\pm8,\pm 5,\pm2, \mp1,\mp4,\mp7\right\rangle$, respectively. We restrict our discussion to a qualitative description of the observed crystal field transitions, as the frequency range of $20-80\cm$ in our far-IR measurements does not allow us to characterize all magnetic-dipole active crystal field transitions and determine the crystal field parameters. For example, the $12.5\cm$ transition observed by inelastic neutron scattering\cite{vajk:87601} is outside of our accessible frequency range.

The magnetic-dipole selection rule for the ($h\bot c$) polarization allows only transitions with $\Delta m=\pm1$. This rule forbids transitions between singlets, but allows transitions between singlets and doublets and between different doublets. Transitions 1 and 2 are strong in the ($h\bot c$) polarization (Fig.~\ref{fig:hperpc}) but are very faint in the ($h||c$) polarization, which suggests that they are allowed in the former, but forbidden in the latter, polarization. Since we observe transition 2 in the highest applied magnetic fields, it must be the transition from the ground state. The final state of the transition could be either a doublet or a singlet. With a magnetic field $B$ applied along the $c$ axis (left panel of Fig.~\ref{fig:hperpc} and its inset) the transition splits linearly, a characteristic of the Zeeman splitting of a doublet. The lower branch of the Zeeman-split transition disappears when the applied field increases beyond 3 T, which is indicative of a thermal depopulation of the upper branch of the Zeeman-split ground state doublet. The slope of the upper branch of transition 2 allows an estimation of the $g$-factor in the effective spin Hamiltonian $\mathcal{H}_{eff}=g\mu_BB\widetilde{S}$, $\widetilde{S}=1/2$, that describes the ground state. Assuming that the final state of the transition is a singlet with energy independent of the applied field (at sufficiently low fields), we get $g_z\approx8.8$. The magnetic field $B$ perpendicular to the $c$ axis produces no Zeeman splitting of the ground state doublet. Therefore, neglecting the mixing of the ground state with higher energy levels at sufficiently low field $B$, we set $g_x=g_y=0$. The upper branch of transition 2 is further split by about $5\cm$ in magnetic fields along the $c$ axis, which we attribute to hyperfine interactions.

\begin{figure}[ht]
\begin{center}
\includegraphics[width=5in]{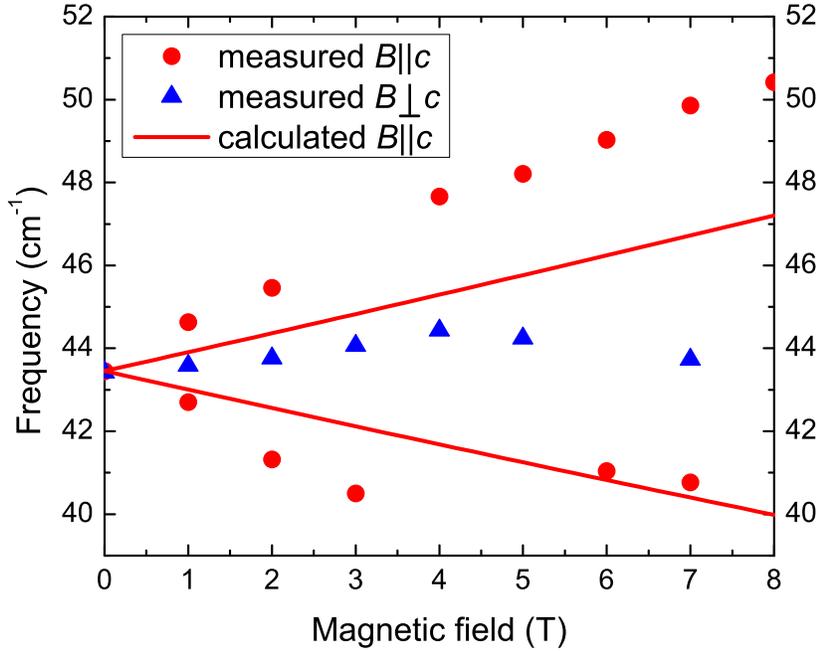}
\caption{\label{fig:afmrcalc}Field dependence of triangular AFMR frequencies at T=10 K. The measured splitting of the resonance by magnetic field is about twice the splitting calculated using the free energy of Eq. (\ref{eq:afmrham}).}
\end{center}
\end{figure}

We now turn to the magnetic response of Mn ions in HMO and assign the sharp absorption line at $43.4\cm$ in the ($h\bot c$) polarization to the antiferromagnetic resonance (AFMR) of the triangular lattice of Mn magnetic moments (transition 5 in Figs.~\ref{fig:lowtspectra},~\ref{fig:hperpc}). This assignment is supported by the observation of the triangular AFMR at the same frequency ($\approx 43\cm$) and light polarization in hexagonal YMnO$_3$\cite{penney:1234}, a related compound with the same Mn magnetic structure and similar exchange and anisotropy parameters\cite{sato:14432}. The AFMR corresponds to the gapped $k=0$ magnon detected by inelastic neutron scattering at $~47\cm$ in HMO\cite{vajk:87601}. The similarity between the $k=0$ magnon frequency and the position of the AFMR in our data further supports the assignment of the resonance. The magnetic field dependence of the AFMR agrees qualitatively with the expected linear splitting induced by the field applied along the $c$ axis ($B||c$) (left panel of Fig.~\ref{fig:hperpc}). A slight increase in the AFMR frequency and no splitting of the resonance are observed in fields up to 5 T applied perpendicular to the $c$ axis ($B\bot c$). In fields of 5 T and higher, the AFMR line is hard to identify, as it becomes a part of the broad absorption band due to the crystal field levels of Ho ions (right panel of Fig.~\ref{fig:hperpc}). The same qualitative behaviour, $i.e.$, linear splitting with $B||c$ and no splitting with $B\bot c$, in fields up to 5.4 T was recorded for the AFMR in YMnO$_3$\cite{penney:1234}.

To quantify the magnetic-field behaviour of the AFMR, we use the following free energy:
\begin{equation}
F=\lambda\sum_{i\neq j}{\bm{M_i\cdot M_j}}+K\sum_i{(M^z_i)^2}-B\sum_i{M^z_i},
\label{eq:afmrham}
\end{equation}
where $M_i$, $i=1,2,3$, are the three sublattice magnetizations within the $a-b$ plane, $\lambda$ describes the antiferromagnetic exchange between the sublattices, and $K$ represents the easy-plane anisotropy. The $B=0$ excitation spectrum of the free energy (\ref{eq:afmrham}) consists of a zero-frequency mode corresponding to global rotations of the sublattices about the $c$ axis and a doubly-degenerate gapped mode whose degeneracy is lifted by the application of $B||c$. The analytical expression for the frequency of the gapped mode in $B||c$ was given by Palme $et$ $al$\cite{palme:873}:
\begin{eqnarray}
\omega^2_{\pm} & = & \frac{ab}{2}-\frac{b(a-b)}{2(a+b)^2}B^2\nonumber\\
 & & \pm \frac{bB}{2(a+b)^2}\sqrt{B^2b(b-2a)+2ab(a+b)^2},
\label{eq:palme}
\end{eqnarray}
where $a=2H_d$, $b=3H_{ex}$, the exchange field $H_{ex}=\lambda M_0$, the anisotropy field $H_d=KM_0$, and $M_0$ is the sublattice magnetization. Equation (\ref{eq:afmrham}) is a molecular-field version of the Hamiltonian used by Vajk $et$ $al$ to describe the spinwave dispersion in HMO\cite{vajk:87601}. In this picture, $H_{ex}=3SJ=127$ T and $H_d=SD=5.75$ T, where $J$ and $D$ are the exchange and anisotropy parameters measured by inelastic neutron scattering and $S$ is the total spin of Mn ions. The value of $H_d$ was lowered slightly to reproduce the zero-field AFMR frequency observed in our far-IR study. The field dependence of the AFMR calculated using Eq. (\ref{eq:palme}) and the $g$-factor $g=2$ is shown in Fig.~\ref{fig:afmrcalc}. The calculated frequencies largely disagree with the measured ones as the measured splitting is about twice the calculated splitting. For comparison, the splitting of the AFMR in YMnO$_3$ is in good agreement with our calculation: the measured splitting of $4.8\cm$ at $\approx5.4$ T ($g=1.9\pm 0.1$)\cite{penney:1234} agrees with the calculated value of $4.9\cm$, as determined from exchange and anisotropy fields measured by inelastic neutron scattering\cite{sato:14432} in YMnO$_3$\cite{footnote1}.

What causes the large discrepancy between the measured and calculated splitting of the AFMR line in HMO? Our calculation describes well the AFMR in YMnO$_3$, a material that is largely similar to HMO, but with one important difference: the Y$^{3+}$ ions are not magnetic. Therefore, we conclude that the magnetic Ho ions in HMO are responsible for the discrepancy. The Mn ions in HMO interact with the surrounding Ho ions via an exchange coupling of the form $J'_{ij}\bm{J^{Ho}_i}\cdot \bm{S^{Mn}_j}$, where $\bm{S^{Mn}}$ and $\bm{J^{Ho}}$ are the angular momenta of Mn and Ho ions. Using the equivalence of matrix elements $g_JJ_z=g_z\widetilde{S}_z$, where $g_J=5/4$ is the Lande factor, $g_z$ and $\widetilde{S}_z$ are the effective g-factor and spin for the Ho ion ground state, we can write the exchange Hamiltonian as\cite{abragam}
$\mathcal{H}_{ex}=\mathcal{J}_x\widetilde{S}_xS^{Mn}_x + \mathcal{J}_y\widetilde{S}_yS^{Mn}_y +
\mathcal{J}_z\widetilde{S}_zS^{Mn}_z,
\label{eq:homnex}$
where $\mathcal{J}_z=(g_z/g_J)J'_{ij}$, etc. We found earlier that $g_z\approx8.8$ and $g_x=g_y=0$, which reduces the Hamiltonian to $\mathcal{H}_{ex} = \mathcal{J}_z\widetilde{S}_zS^{Mn}_z$. In magnetic field $B$ applied along the $c$ axis, we can replace $\widetilde{S}_z$ by its thermal average $\left\langle \widetilde S_z\right\rangle=\chi B/g_z\mu_B$ and write $\mathcal{H}_{ex} =(\mathcal{J}_z\chi)/(g_z\mu_B)BS^{Mn}_z$, where $\chi$ is the magnetic susceptibility of Ho ions. The last expression shows that the Ho-Mn (HM) exchange interaction is equivalent to an effective magnetic field acting on Mn spins. To describe the effect of the HM exchange on the AFMR, we add the term
\begin{equation}
F_{HM}=\frac{\mathcal{J}_z\chi N}{g_zg\mu_B^2}B\sum_{i}M_i^z=\lambda_{HM}B\sum_{i}M_i^z,
\label{eq:homnfe}
\end{equation}
where $N$=6 is the number of Ho ions neighboring a Mn ion, to the free energy of the Eq. (\ref{eq:afmrham}). The total magnetic field acting on Mn ions is then the sum of the external static field and the effective field of Ho ions, $B^{tot}=(1-\lambda_{HM})B$, and can be larger or smaller than the external field, thus enhancing or mitigating its effect on the AFMR. Comparing the measured AFMR splitting and the splitting calculated using Eq. (\ref{eq:afmrham}), we find that $\lambda_{HM}=-1$, which corresponds to ferromagnetic HM coupling.

We can account for the two kinds of Ho lattice sites (Ho(1) and Ho(2)) by writing $\mathcal{J}_z=(2\mathcal{J}^1_z + 4\mathcal{J}^2_z)/N$, where $\mathcal{J}^1_z$ and $\mathcal{J}^2_z$ are the HM exchange constants for each site. Each Mn ion has two Ho(1) ions and four Ho(2) ions as Ho nearest neighbors. The two Ho lattice sites are distinct because of the ferroelectric displacements. Such displacements are significantly smaller that interionic distances in HoMnO$_3$, and the approximation $\mathcal{J}^1_z\approx\mathcal{J}^2_z\approx\mathcal{J}_z$ is expected to be a good one. Otherwise, we must use the above combination of $\mathcal{J}^1_z$ and $\mathcal{J}^2_z$ in the definition of $\lambda_{HM}$ in Eq. (\ref{eq:homnfe}).

Before we conclude, we consider the possible effect of phonon coupling on the frequencies of the studied magnetic excitations. The most prominent signature of such coupling is the anti-crossing of the coupled magnetic and lattice modes. Such anti-crossing is expected to be strongest when the frequencies of the two modes are similar, as shown on the right panel of Fig.~\ref{fig:hperpc}, where the strong magnetic absorption line overlaps at high magnetic field with the non-magnetic absorption at 70$\cm$. At all magnetic fields, we see no measurable changes to the 70$\cm$ excitation. Thus, we conclude that we detect no measurable coupling between magnetic and lattice vibrations.

The measured ferromagnetic HM exchange plays a significant role in all aspects of the physics of HoMnO$_3$. It causes the electric-field induced ferromagnetic ordering of Ho magnetic moments along the $c$ axis\cite{lottermoser:541} and may also be responsible for the multitude of low-temperature (T$<$8 K) magnetic phases and the two Mn spin reorientation transitions in HMO\cite{vajk:87601}. The reentrant phase that is induced in HMO by magnetic field displays a strong magnetodielectric effect and was explained by the formation of magnetic domain walls during spin reorientation, which reduces the local magnetic symmetry and allows the coupling between magnetic and ferroelectric polarizations\cite{lorenz:87204, lottermoser:220407}. Theoretical descriptions of such magnetoelectricity must include domain wall contributions to the free energy from lattice distortions and magnetic superexchange interactions. Such descriptions are incomplete without the consideration of the magnetic exchange between the Ho and Mn ions. Our measurement of the ferromagnetic HM exchange provides an important contribution to the modeling of the interplay between magnetism and ferroelectricity in HoMnO$_3$.

We thank Darryl Smith for useful theoretical discussions and Jonathan Gigax for the simulation of the Ho crystal field eigenstates. The work at LANL was supported by the LDRD program and the Center for Integrated Nanotechnologies. The work at UCSD was supported by NSF DMR 0705171.


\begin{thebibliography}{22}
\bibitem[{foo()}]{footnote1}
\bibinfo{note}{Sato $et$ $al.$\cite{sato:14432} find that the chemical
  triangular lattice of Mn ions is slightly trimerized with exchange strength
  $J_1$ between the nearest neighbors belonging to the same trimer and exchange
  strength $J_2$ between the ions belonging to neighboring trimers. The zero
  wave-vector magnon spectrum of the trimerized chemical lattice is calculated
  using Eq. (\ref{eq:palme}) and the substitution $J\rightarrow (J_1 +
  2J_2)/3$.}
\end{thebibliography}

\end{document}